\documentclass[letterpaper,11pt]{article}

\usepackage{amsmath,amssymb,amstext,mathrsfs}
\usepackage{slashed}

\usepackage{graphicx}

\usepackage{tikz}
\usepackage{pgfplots}
\pgfplotsset{compat=1.18}
\usepgfplotslibrary{fillbetween}
\usetikzlibrary{patterns}

\usepackage{cite}

\usepackage{hyperref}

\newcommand{\nn}{\nonumber}
\allowdisplaybreaks

\makeatletter
	\@addtoreset{equation}{section}
	\makeatother

\begin{document}
		\begin{titlepage}
			
			\bigskip
			\begin{center}
				{\LARGE\bfseries  Chern-Simons-like formulation of 3D MMG-like massive gravity models}
				\\[10mm]
				\textbf{B\"u\c{s}ra Dedeo\u{g}lu
				$^1$, Mehmet Ozkan
				$^2$ and \"Ozg\"ur Sar{\i}o\u{g}lu
				$^1$}\\[5mm]
				\vskip 25pt
				
				{\em  \hskip -.1truecm $^1$ Department of Physics, Faculty of Arts and Sciences,
					Middle East Technical University, 06800, Ankara, T\"urkiye  \vskip 10pt }
				
				{\em  \hskip -.1truecm $^2$Department of Physics, İstanbul Technical University,  \\
					Maslak, 34469, İstanbul, T\"urkiye  \vskip 10pt }

				{email: {\tt busrad@metu.edu.tr, ozkanmehm@itu.edu.tr, sarioglu@metu.edu.tr}}
				
			\end{center}
			
			\vspace{3ex}

			\begin{center}
				{\bfseries Abstract}
			\end{center}
			\begin{quotation} \noindent

We investigate the Chern–Simons-like formulation of 3D MMG-like massive gravity models that are
 ``third-way consistent". Building on previous work on exotic massive gravities, we analyze a class of 
 MMG-like theories characterized by a specific parity structure and an auxiliary field hierarchy. Focusing 
 on the simplest non-trivial case, we solve the full set of field equations, determine the AdS background 
 solutions, compute the central charges of the dual CFT, and perform a linearized analysis to obtain the 
 mass spectrum. Along the chiral line, the linearized mass operator develops a rank-2 Jordan block, 
 signaling logarithmic behavior of massive modes in the dual two-dimensional CFT. At a special 
 degenerate point, this structure is enhanced to a rank-3 Jordan block, giving rise to two logarithmic 
 partners and an ultra-logarithmic sector in the boundary theory.

            \end{quotation}
			
			\vfill
	
			\flushleft{\today}
		\end{titlepage}
		
		\tableofcontents
			
\bigskip

\section{Introduction}
\paragraph{}
In three dimensions, the main motivation for considering modified gravity is to generate a local propagating degree of freedom, and this is also where most of the difficulties lie. The (cosmological) Einstein gravity is essentially topological: it has no local graviton, and attempts to introduce one, typically by adding higher-derivative terms, trigger a whole chain of pathologies. Ghosts appear through wrong-sign kinetic terms, tachyonic or non-unitary modes show up in the bulk spectrum, and bulk unitarity often clashes with the positivity of the Brown–Henneaux central charges at the boundary \cite{Merbis:2014vja}. From a technical viewpoint, these pathologies arise because each consistency requirement imposes a constraint on the parameter space of the theory. The central challenge of three-dimensional modified gravity is therefore to introduce as many independent free parameters as possible, so that the number of adjustable parameters matches the number of constraints.

To be concrete, consider the Topological Massive Gravity (TMG), which is the simplest modification to Einstein's gravity in three dimensions with field equation \cite{Deser:1982vy,Deser:1981wh} 
\begin{equation}
    0 = \sigma G_{\mu\nu} + \Lambda_0 g_{\mu\nu} + \frac{1}{\mu} C_{\mu\nu} \,,
    \label{TMG-EOM}
\end{equation}
where $\sigma = \pm 1$. Here $C_{\mu\nu}$ is the Cotton tensor, and the model has three free parameters ($\sigma, \Lambda_0, \mu)$. On the other hand, TMG contains a single propagating degree of freedom, leading to the following constraints for the perturbative unitarity on the parameter space: positivity of the Fierz-Pauli mass squared, the sign of the kinetic term of the Fierz-Pauli action, and the positivity of the central charges. Although it seems like the theory has as many free parameters as the number of constraints, in fact it only has two dimensionless parameters, $\sigma, \mu \ell$ (where $\ell$ is the AdS$_3$ radius), and all constraints are expressed in terms of these two parameters. As a result, the number of independent parameters is actually smaller than the number of constraints.  This picture is not improved by adding further higher-derivative terms or by mixing models at different derivative orders \cite{Afshar_2014}.

The situation improves if one relaxes the requirement that the gravity model be described by a single metric Lagrangian. In that case, bi-gravity models provide a natural generalization: by introducing an additional metric (or dreibein) and the corresponding interactions, they realize a larger parameter space and have been shown to admit regions where all consistency conditions can be simultaneously satisfied \cite{Afshar_2014,Bergshoeff:2013xma,Ozkan:2019iga,Sevim:2019scg}. A conceptually different alternative is the so-called “third way” to three-dimensional gravity. Here, the equations of motion can still be written in terms of a single metric, but the divergence of the field equation is proportional to the field equation itself \cite{Bergshoeff:2015zga}. This weakened, on-shell notion of covariant conservation is what makes these theories consistent, and it leads to a distinctive feature: third-way models that cannot be derived from any standard covariant single-metric Lagrangian. To make this more concrete, let us recall that the TMG field equation \eqref{TMG-EOM} can be deformed into the Minimal Massive Gravity (MMG) equation \cite{MMG}, which enlarges the parameter space without introducing an additional degree of freedom
\begin{equation}
0 = \sigma G_{\mu\nu} + \Lambda_0 g_{\mu\nu} + \frac{1}{\mu} C_{\mu\nu} + \frac{1}{2m^2} \epsilon_{\mu}{}^{\rho \sigma} \epsilon_{\nu}{}^{\lambda \tau} S_{\rho\lambda} S_{\sigma \tau} \,,
\label{MMG-EOM}
\end{equation}
where $S_{\mu\nu}$ is the three-dimensional Schouten tensor. As mentioned, the divergence of this field equation refers back to itself, leading to the third-way, on-shell notion of covariant conservation of the metric field equation \cite{MMG}. At the same time, the presence of the new parameter $m^2$ enlarges the parameter space sufficiently to resolve the unitarity problems that afflict TMG.

The introduction of MMG naturally leads to two closely related questions. First, to what extent does the third-way idea generalize to other massive gravity models that previously failed to meet all consistency requirements? For instance, can one “upgrade’’ New Massive Gravity (NMG) \cite{Bergshoeff:2009hq} or General Massive Gravity (GMG) \cite{Bergshoeff:2009aq} to third-way theories, thereby enlarging their parameter space and restoring unitarity, without introducing additional propagating degrees of freedom? Second, from a more structural point of view, the third-way mechanism is highly selective, so one would like a systematic procedure for constructing third-way Lagrangians, within a Chern–Simons–like formulation \cite{Bergshoeff:2014bia}, that guarantees the on-shell conservation property. In an attempt to enlarge the parameter space of NMG and GMG, the Exotic Massive Gravity (EMG) and Exotic General Massive Gravity (EGMG) models were introduced as third-way extensions \cite{Ozkan-exotic}. These theories have the same perturbative spectrum as NMG and GMG, respectively, while satisfying the on-shell covariant conservation property. However, unlike MMG, the third-way generalizations of NMG and GMG do not enlarge the space of independent parameters, and these models still fail to satisfy the bulk and boundary unitarity constraints simultaneously \cite{Ozkan-exotic}. 

The reason MMG is such a unique theory can be seen by considering the generic form of a third-way equation of motion \cite{Ozkan-exotic}
\begin{eqnarray} \label{3rdWayEOM}
   E_{\mu\nu} := \Lambda_0 g_{\mu\nu} + G_{\mu\nu} + \mathcal{G}_{\mu\nu} +\mathcal{H}_{\mu\nu} + \mathcal{L}_{\mu\nu} = 0 \,, 
\end{eqnarray}
where $G_{\mu\nu}$ is the Einstein tensor, and $\mathcal{H}_{\mu\nu}$ and $\mathcal{L}_{\mu\nu}$ are defined as
		\begin{align}
			\mathcal{H}_{\mu\nu} &:= \epsilon_\mu{}^{\rho\sigma} \nabla_{\rho} \mathcal{S}_{\nu\sigma} \,, & \mathcal{L}_{\mu\nu} & := \frac12 \epsilon_\mu{}^{\rho\sigma} \epsilon_\nu{}^{\lambda\tau} \mathcal{S}_{\rho\lambda} \mathcal{S}_{\sigma\tau} \,.
		\end{align}
		Here, $\mathcal{S}_{\mu\nu}$ is a Schouten-like tensor that is obtained from an Einstein-like tensor $\mathcal{G}_{\mu\nu}$ with $\nabla^{\mu} \mathcal{G}_{\mu\nu} = 0$ as
		\begin{eqnarray}
			\mathcal{S}_{\mu\nu} := \mathcal{G}_{\mu\nu} - \frac12 g_{\mu\nu} g^{\rho\lambda} \mathcal{G}_{\rho\lambda} \,. 
		\end{eqnarray}
        The Einstein-like tensor $\mathcal{G}_{\mu\nu}$ in \eqref{3rdWayEOM} does not play any direct role in ensuring that the covariant divergence of the field equation is proportional to $E_{\mu\nu}$, but it can nevertheless be included consistently. An extra care is required for the dimensional coefficients multiplying the various contributions in \eqref{3rdWayEOM}, in particular those in front of $\mathcal{H}_{\mu\nu}$ and $\mathcal{L}_{\mu\nu}$. The divergence condition implies that if the coefficient of $\mathcal{H}_{\mu\nu}$ is $\alpha$, then the coefficient of $\mathcal{L}_{\mu\nu}$ must be $\alpha^2$. Hence these two terms do not introduce two independent parameters into the theory. An exception arises when $\mathcal{H}_{\mu\nu}$ is the Cotton tensor, $C_{\mu\nu}$, corresponding to choosing the Einstein-like tensor to be the Einstein tensor $\mathcal{G}_{\mu\nu} = G_{\mu\nu}$. In this case the coefficients of $\mathcal{H}_{\mu\nu}$ and $\mathcal{L}_{\mu\nu}$ can be treated independently since $\nabla^\mu C_{\mu\nu} = 0$, and one may introduce a free coefficient both for $\mathcal{H}_{\mu\nu}$ and $\mathcal{L}_{\mu\nu}$, increasing the number of free parameters and matching the number of constraints from the bulk and boundary unitarity. This special choice precisely corresponds to MMG \cite{MMG}.

        While it is not straightforward, if possible, to increase the number of free parameters in third-way models beyond MMG, it is nevertheless important to understand their most general structure, since third-way models are inherently difficult to construct as they do not descend from an invariant single-metric action. In our previous work, we achieved such a characterization for the “exotic’’ class of models \cite{my-precious}, which exhibit parity-odd behavior and can be viewed as exotic cousins of parity-even massive gravities. From the point of view of derivative counting, these exotic theories have an even number of derivatives as the highest derivative order acting on the metric in the equation of motion. The purpose of the present work is to complete the analysis of the third way to three-dimensional gravity by establishing the structure of MMG-like models in which the highest derivative order acting on the metric is odd. In Section \ref{cs-formulation}, we set the stage and first briefly recall the torsional construction of third-way models. We, then, construct the Chern–Simons–like formulation of MMG-like models of third-way to three-dimensional gravity and discuss its properties assuming an AdS$_3$ background solution. Building on this, in Section \ref{mmg-like models}, we provide an in-depth analysis of the simplest model beyond MMG, including its central charges, the corresponding unitarity conditions and the chiral limit. Our conclusions are given in Section~\ref{conclusions}.

        \section{Chern-Simons-like formulation of MMG-like models} \label{cs-formulation}
\paragraph{}
In this section we briefly recall the basics of Chern--Simons-like (CS-like) theories of gravity, which will set the stage for introducing MMG-like third-way models of massive gravity. CS-like models of massive gravity are three-dimensional theories that can be written in the form \cite{Bergshoeff:2014bia}
\begin{equation}
    L = \frac{1}{2}\, g_{rs}\, a^r \cdot d a^s + \frac{1}{6}\, f_{rst}\, a^r \cdot a^s \times a^t \,,
\end{equation}
where $a^{r}$ are Lorentz-vector valued one-form fields. These fields include the dreibein $e^a$, the dual spin connection $\omega^a$, as well as a finite number of auxiliary one-forms, collectively labeled by the flavour index $r$. The constants $g_{rs}$ define a ``metric" on flavour space, while $f_{rst}$ is a totally symmetric flavour tensor. Here, the dot and cross denote contraction with the Lorentz metric and the Levi-Civita symbol, respectively:
\begin{equation}
    x \cdot y := \eta_{ab} x^a y^b \,, 
    \qquad
    (x \times y)^a := \epsilon^{a}{}_{bc} x^b y^c \,.
\end{equation}
In this formulation, the equations of motion are solved algebraically for the spin connection and the auxiliary fields in terms of the dreibein (and its derivatives), so that the dreibein remains the only independent dynamical field and one recovers a higher-derivative massive theory of gravity at the metric level. The CS-like formulation also assumes definite parity assignments for the flavour fields $a^r$. In particular, we take the fundamental fields, namely the dreibein and the (undualized) spin connection, to be parity even, so that the dual spin connection $\omega^a$ is assigned odd parity. Based on these assignments, we will denote parity-even auxiliary one-forms by $h_{(n)}$ and parity-odd auxiliary one-forms by $f_{(n)}$. In order to discuss how to construct MMG-like models in a CS-like formulation, it is useful to start from their dynamical equation of motion. After setting all dimensional constants to unity, the third-way equation can be written schematically as
\begin{eqnarray} \label{third-way-eqn-A}
    0
    = R(\omega)
    + \frac{1}{2}\, e \times e
    + e \times \mathcal{A}
    + D(\omega) \mathcal{A}
    + \frac{1}{2}\, \mathcal{A} \times \mathcal{A} \,,
\end{eqnarray}
where $\mathcal{A}$ is a Lorentz-valued one-form corresponding to a Schouten-like tensor which, in the CS-like formulation, is expressed in terms of the auxiliary fields $h_{(n)}$, $f_{(n)}$ and their combinations. Here $R(\omega)$ denotes the curvature two-form of the dual spin connection $\omega$, and $D(\omega)$ is the associated Lorentz-covariant exterior derivative. If we now interpret $\mathcal{A}$ as the torsion, we may define a torsional dual spin connection
\begin{equation}
    \Omega := \omega + \mathcal{A} \,.
\end{equation}
In terms of $\Omega$, the third-way equation \eqref{third-way-eqn-A} can be rewritten as
\begin{eqnarray} \label{ads-eqn}
    0 = R(\Omega) + \frac{1}{2\ell^{2}}\, e \times e \,,
\end{eqnarray}
where we have reinstated the AdS radius $\ell$. Thus, the third-way equation of motion takes the form of a torsionful AdS$_3$ Einstein equation for the connection $\Omega$.

As discussed in \cite{my-precious}, in the case of exotic and exotic general theories the CS-like formulation is obtained by choosing
\begin{align}
  \text{Exotic} &:   \mathcal{A} = \sum_{n \geq 1} a_{n}\, f_{(n)} \,,\nonumber\\
  \text{Exotic General}& :   \mathcal{A} = \sum_{n \geq 1} a_{n}\, f_{(n)} + e  \,.
\end{align}
In the present work, we extend these constructions to include MMG-like models. To construct a torsional MMG-like Lagrangian, we choose the highest-order parity-even auxiliary field $h_{(N)}$ to play the role of torsion. This field is paired with the highest-order parity-odd auxiliary field $f_{(N-1)}$ in such a way that the variation of the Lagrangian with respect to $f_{(N-1)}$ yields the torsion constraint. Consequently, $f_{(N-1)}$ appears in the action only linearly, and the resulting third-way equation \eqref{third-way-eqn-A} can involve $f$-fields only up to $f_{(N-2)}$. Based on this structure of the MMG-like models, it is natural to interpret $\mathcal{A}$ as a torsion one-form given by
\begin{eqnarray} \label{a-parity mixed}
    \mathcal{A}=e+\sum_{n=1}^{N-2}a_nf_{(n)}+\sum_{n=1}^N b_nh_{(n)}\,.
\end{eqnarray} 
We start our investigation by first presenting the $\mathcal{A} \propto h_{(N)}$ case in detail. We also ignore the dreibein $e$ in \eqref{a-parity mixed}, since field redefinitions can redistribute the $e$-dependence between the Schouten-like tensor and the auxiliary fields.
The Lagrangian of the model is obtained by choosing $h_{(N)}$ as torsion,
\begin{align} \label{lagrangian-MMG-bn=0}
   \mathcal{L}^N =\;& f_{(N-1)} \cdot D(\Omega)e
   + h_{(N-1)} \cdot R(\Omega)
   + e \cdot f_{(N-1)} \times h_{(N)}
   + \frac{1}{2} b_N\, e \cdot e \times h_{(N)}   \nn \\
   & + \frac{1}{2} \sum_{n=1}^{N-1} e \cdot h_{(n)} \times h_{(N-n)}
   + \frac{1}{2} \sum_{n=1}^{N-2} \sum_{m=1}^{n-1}
      h_{(m)} \cdot f_{(n-m)} \times f_{(N-n-1)}
   + h_{(N)} \cdot D(\Omega) h_{(N-1)} \nn \\
   & + \sum_{n=1}^{N-2} f_{(N-n-1)} \cdot D(\Omega) h_{(n)}
   + \frac{\gamma}{2}\, \Omega \cdot
      \left(d \Omega + \frac{1}{3}\, \Omega \times \Omega \right)
   + \frac{1}{2} \sum_{n=1}^{N-2} e \cdot f_{(n)} \times f_{(N-n-1)} \nn \\
   & + \sum_{n=1}^{N-2} h_{(n)} \cdot h_{(N)} \times f_{(N-n-1)}
   + \frac{1}{6} \sum_{n=1}^{N-2} \sum_{m=1}^{n}
      h_{(m)} \cdot h_{(n-m+1)} \times h_{(N-n-1)}  \nn \\
   & + \frac{1}{2}\, h_{(N-1)} \cdot h_{(N)} \times h_{(N)} \,.
\end{align}
The corresponding field equations are
\begin{align}
\delta f_{(N-1)} \, :&\quad 0 = D(\Omega) e + e \times h_{(N)} \,, \nn \\
\delta f_{(N-n-1)} \, :&\quad
  0 = D(\Omega) h_{(n)} + h_{(n)} \times h_{(N)}
    + e \times f_{(n)}
    + \sum_{m=1}^{n-1} h_{(m)} \times f_{(n-m)} \,,
    \quad n \leq N-2 \,, \nn \\
\delta e \, :&\quad
  0 = D(\Omega) f_{(N-1)}
    + f_{(N-1)} \times h_{(N)}
    + b_N\, e \times h_{(N)}
    + \frac{1}{2} \sum_{n=1}^{N-1} h_{(n)} \times h_{(N-n)} \nn \\
  &\qquad 
    + \frac{1}{2} \sum_{n=1}^{N-2} f_{(n)} \times f_{(N-n-1)} \,, \nn \\
\delta \Omega \, :&\quad
  0 = D(\Omega) h_{(N-1)}
    + h_{(N-1)} \times h_{(N)}
    + e \times f_{(N-1)}
    + \sum_{n=1}^{N-2} h_{(n)} \times f_{(N-n-1)}
    \nn\\
    & \qquad + \gamma\, R(\Omega) \,, \nn \\
\delta h_{(N-n-1)} \, :&\quad
  0 = D(\Omega) f_{(n)}
    + f_{(n)} \times h_{(N)}
    + e \times h_{(n+1)}
    + \frac{1}{2} \sum_{m=1}^{n} h_{(m)} \times h_{(n-m+1)} \nn \\
  &\qquad
    + \frac{1}{2} \sum_{m=1}^{n-1} f_{(n-m)} \times f_{(m)} \,,
    \qquad n \leq N-2 \,, \nn \\
\delta h_{(N-1)} \, :&\quad
  0 = D(\Omega) h_{(N)}
    + R(\Omega)
    + \frac{1}{2}\, h_{(N)} \times h_{(N)}
    + e \times h_{(1)}
    + \frac{1}{2} b_{N-1}\, e \times e \,, \nn \\
\delta h_{(N)} \, :&\quad
  0 = D(\Omega) h_{(N-1)}
    + h_{(N-1)} \times h_{(N)}
    + e \times f_{(N-1)}
    + \sum_{n=1}^{N-2} h_{(n)} \times f_{(N-n-1)}
   \nn\\
   & \qquad + \frac{1}{2} b_N\, e \times e \,.
\end{align}
First, note that subtracting the $\delta h_{(N)}$ equation from the $\delta \Omega$ equation gives
\begin{equation}
    0
    = \, \delta \Omega - \delta h_{(N)} = \, \gamma\, R(\Omega) - \frac{1}{2} b_N\, e \times e \,.
\end{equation}
The AdS condition then fixes the coefficient of the Chern--Simons term, as in the exotic case, to
\begin{equation}
    \gamma = - \ell^2 b_N \,.
\end{equation}
To study the background solutions, we perform the transformation $\Omega = \omega - h_{(N)}$
after which the torsion-free field equations take the form
\begin{align}
\delta f_{(N-1)} \, :&\quad 0 = D(\omega) e \,, \nn \\
\delta h_{(N-1)} \, :&\quad 0 = R(\omega) + e \times h_{(1)} \,, \nn \\
\delta f_{(N-n-1)} \, :&\quad
  0 = D(\omega) h_{(n)}
    + e \times f_{(n)}
    + \sum_{m=1}^{n-1} h_{(m)} \times f_{(n-m)} \,,
    \quad n \leq N-2 \,, \nn \\
\delta h_{(N-n-1)} \, :&\quad
  0 = D(\omega) f_{(n)}
    + e \times h_{(n+1)}
    + \frac{1}{2} \sum_{m=1}^{n} h_{(m)} \times h_{(n-m+1)}
    \,, \nn \\
  &\qquad + \frac{1}{2} \sum_{m=1}^{n-1} f_{(m)} \times f_{(n-m)} \,, \qquad  n \leq N-2 \,, \nn \\
\delta h_{(N)} \, :&\quad
  0 = D(\omega) h_{(N-1)}
    + e \times f_{(N-1)}
    + \frac{1}{2} b_N\, e \times e
    + \sum_{n=1}^{N-2} h_{(n)} \times f_{(N-n-1)} \,, \nn \\
\delta e \, :&\quad
  0 = D(\omega) f_{(N-1)}
    + b_N\, e \times h_{(N)}
    + \frac{1}{2} \sum_{n=1}^{N-2} f_{(n)} \times f_{(N-n-1)}
   \nn\\
   & \qquad + \frac{1}{2} \sum_{n=1}^{N-1} h_{(n)} \times h_{(N-n)} \,.    
\end{align}
As is evident from the $\delta e$ equation, the model can be extended to include all $b_n$ in the definition of $\mathcal{A}$ with the following simple modification:
\begin{eqnarray}
    \Tilde{\mathcal{L}}
    = \mathcal{L}^N
    + \frac{1}{2} \sum_{n=1}^{N-1} b_n\, e \cdot e \times h_{(n)} \,.
    \label{ExoticAction2}
\end{eqnarray}
With this modification, the $\delta e$ equation becomes
\begin{eqnarray}
    \delta e \, :&\quad
    0 &= D(\Omega) f_{(N-1)}
      + f_{(N-1)} \times h_{(N)}
      + \sum_{n=1}^{N-1} b_n\, e \times h_{(n)}
      + b_N\, e \times h_{(N)} \nn \\
    &&\quad
      + \frac{1}{2} \sum_{n=1}^{N-1} h_{(n)} \times h_{(N-n)}
      + \frac{1}{2} \sum_{n=1}^{N-2} f_{(n)} \times f_{(N-n-1)} \,.
\end{eqnarray}
The metric equation is then of the generic third-way form \eqref{third-way-eqn-A} with $\mathcal{A} = -h_{(N)}$,
\begin{eqnarray}
   0 = R(\omega)
     - D(\omega) h_{(N)}
     + \frac{1}{2}\, h_{(N)} \times h_{(N)}
     + \frac{1}{2\ell^2}\, e \times e \,,
\end{eqnarray}
where the AdS condition still fixes the coefficient of the Lorentz–Chern–Simons term $\gamma$ in \eqref{lagrangian-MMG-bn=0} to be
\begin{equation}
    \gamma = - \ell^2 b_N \,.
\end{equation}
We may, however, include an Einstein-like tensor, which can be achieved by adding a cosmological constant term to the action,
\begin{eqnarray}
    \mathcal{L}
    = \mathcal{L}^N
    + \frac12 \sum_{n=1}^{N-1} b_n\, e \cdot e \times h_{(n)}
    + \frac16 \Lambda_0 \, e \cdot e \times e \,.
    \label{ExoticAction4}
\end{eqnarray}
For a maximally symmetric AdS background, we assume that the auxiliary fields are proportional to the dreibein, i.e.
\begin{equation}
    (\bar{h}_{(n)}, \bar{f}_{(n)}) = (c_n \bar{e}, \beta_n \bar{e}) \,.
\end{equation}
This ansatz allows us to determine the coefficient $c_1$ from the $\delta h_{(N-1)}$ equation as
\begin{eqnarray}
    c_1 = \frac{1}{2}\left( \frac{1}{\ell^2} - b_{N-1} \right)\,.
\end{eqnarray}
Furthermore, the $\delta f_{(N-n-1)}$ equations fix $\beta_n = 0$ for $n \neq N-1$, while the $\delta h_{(N-n-1)}$ and $\delta e$ equations determine the remaining $c_n$ through
\begin{align} \label{bc-MMG-bn=0}
    c_{n+1} &= - \frac{1}{2} \left( b_{N-n-1}
                 + \sum_{m=1}^n c_m c_{n-m+1} \right) \,, 
                 \qquad n \leq N-2\,, \nn \\
    c_N &= - \frac{1}{2 b_N} \left( \sum_{n=1}^{N-1} ( 2 b_n + c_{N-n}) c_n
                 + \Lambda_0 \right)\,.
\end{align}
The field equations, together with the expression for $c_1$, imply that $c_N$ is identically zero. Consequently, the bare cosmological constant $\Lambda_0$ can be written in terms of the parameters $c_n$ as
\begin{eqnarray}
    \Lambda_0 = - \sum_{n=1}^{N-1} c_n \bigl( 2 b_n + c_{N-n} \bigr)\,.
\end{eqnarray}
For models with vanishing bare cosmological constant, $\Lambda_0 = 0$, this relation imposes the constraint
\begin{eqnarray}
    c_n = -2 b_{N-n} \,, \qquad n=1, 2, \dots, N-1 \,.
\end{eqnarray}
Moreover, the torsion-free form of the $\delta \Omega$ field equation shows that the only non-vanishing background value among the $\beta_n$ is
\begin{eqnarray}
    \beta_{N-1} = - \frac{1}{2} b_N \,.
\end{eqnarray}

In the next section we turn to the case $N=2$, which provides the simplest non-trivial example within the MMG-like family of models. Starting from the torsional Lagrangian, we systematically derive the full set of field equations and determine the explicit form of the auxiliary fields by solving the torsion and curvature constraints. This yields the corresponding torsion-free formulation and allows us to evaluate the AdS background solutions. To assess the physical consistency of the $N=2$ model, it is essential to examine the conditions required for the absence of tachyons and ghost states. The no-tachyon condition ensures that all propagating modes have non-negative mass-squared in the AdS background, while the no-ghost condition guarantees the positivity of the kinetic terms after diagonalizing the quadratic action. Both requirements impose non-trivial restrictions on the parameter space. In what follows, we provide a detailed analysis of these conditions and discuss the resulting physically admissible regions for this model.

\section{\texorpdfstring{$N=2$}{Lg} MMG-like models}
\label{mmg-like models}

The most general $N=2$ MMG-like model is described by the Lagrangian
\begin{align} \label{lagrangian}
    \mathcal{L} =\;&
    f_{(1)} \cdot D(\Omega)e
    + e \cdot h_{(2)} \times f_{(1)}
    + h_{(2)} \cdot D(\Omega)h_{(1)}
    + \frac{1}{2}\, h_{(1)} \cdot h_{(2)} \times h_{(2)}
    + h_{(1)} \cdot R(\Omega) \nn \\
    & + \frac{1}{2}\, e \cdot h_{(1)} \times h_{(1)}
    + \frac{\Lambda_0}{6}\, e \cdot e \times e
    + \frac{\rho}{2\ell^2}\, e \cdot e \times h_{(1)}
    - \frac{\gamma}{6\ell^3}\, e \cdot e \times h_{(2)} \nn \\
    & + \frac{\gamma}{6\ell} \left(
          \Omega \cdot d\Omega
          + \frac{1}{3}\, \Omega \cdot \Omega \times \Omega
        \right) \,.
\end{align}
The parameter $\rho$ is dimensionless, whereas $\Lambda_0$ has mass dimension $3$, and $\gamma$ has mass dimension $1$. Consequently, the overall Lagrangian needs to be multiplied by a constant of mass dimension $-1$ in order to have the correct physical dimensions; however, this overall factor does not affect the field equations and will be omitted in what follows. The equations of motion are, therefore, given by
\begin{align}
    \delta f_{(1)} \, :&\quad
    0 = D(\Omega)e + e \times h_{(2)} \,, \nn \\
    \delta h_{(2)} \, :&\quad
    0 = D(\Omega)h_{(1)}
        + h_{(1)} \times h_{(2)}
        + e \times f_{(1)}
        - \frac{\gamma}{6\ell^3}\, e \times e \,, \nn \\
    \delta h_{(1)} \, :&\quad
    0 = D(\Omega)h_{(2)}
        + \frac{1}{2}\, h_{(2)} \times h_{(2)}
        + R(\Omega)
        + e \times h_{(1)}
        + \frac{\rho}{2\ell^2}\, e \times e \,, \nn \\
    \delta e \, :&\quad
    0 = D(\Omega)f_{(1)}
        + f_{(1)} \times h_{(2)}
        + \frac{1}{2}\, h_{(1)} \times h_{(1)}
        + \frac{\Lambda_0}{2}\, e \times e
        + \frac{\rho}{\ell^2}\, e \times h_{(1)}
        - \frac{\gamma}{3\ell^3}\, e \times h_{(2)} \,, \nn \\
    \delta \Omega \, :&\quad
    0 = \frac{\gamma}{3\ell}\, R(\Omega)
        + D(\Omega)h_{(1)}
        + h_{(1)} \times h_{(2)}
        + e \times f_{(1)} \,.
\end{align}
These equations are of the standard first-order form that allows one to solve for $f_{(1)}$, $h_{(1)}$ and $h_{(2)}$ in the presence of torsion. In particular, we can introduce a torsion-free spin connection by defining
\begin{equation}
    \omega = \Omega + h_{(2)} \,,
\end{equation}
which is equivalent to $\Omega = \omega - h_{(2)}$. Combining the $\delta \Omega$ and $\delta h_{(2)}$ equations yields the metric equation
\begin{eqnarray} \label{metric eqn}
    0 = R(\Omega) + \frac{1}{2\ell^2}\, e \times e \,.
\end{eqnarray}
The torsion-free equations of motion, expressed in terms of $\omega$, are
\begin{align}
    \delta f_{(1)} \, :&\quad
    0 = D(\omega)e \,, \nn \\
    \delta h_{(1)} \, :&\quad
    0 = R(\omega)
        + e \times h_{(1)}
        + \frac{\rho}{2\ell^2}\, e \times e \,, \nn \\
    \delta \omega \, :&\quad
    0 = D(\omega)h_{(1)}
        + e \times f_{(1)}
        + \frac{\gamma}{3\ell} \bigl(
            R(\omega) - D(\omega)h_{(2)}
            + \tfrac12 h_{(2)} \times h_{(2)}
          \bigr) \,, \nn \\
    \delta e \, :&\quad
    0 = D(\omega)f_{(1)}
        + \frac{1}{2}\, h_{(1)} \times h_{(1)}
        + \frac{\Lambda_0}{2}\, e \times e
        + \frac{\rho}{\ell^2}\, e \times h_{(1)}
        - \frac{\gamma}{3\ell^3}\, e \times h_{(2)} \,, \nn \\
    \delta h_{(2)} \, :&\quad
    0 = R(\omega)
        - D(\omega)h_{(2)}
        + \frac{1}{2}\, h_{(2)} \times h_{(2)}
        + \frac{1}{2\ell^2}\, e \times e \,.
\end{align}
It is convenient to perform the following simultaneous field redefinitions in order to determine the explicit form of the auxiliary fields:
\begin{align} \label{hftoHF}
    h_{(1)} &\;\to\; H_{(1)} - \frac{\rho}{2 \ell^2}\, e \,, &
    h_{(2)} &\;\to\; \frac{3\rho \ell}{2\gamma}\, H_{(1)}
                     - H_{(2)}
                     + \left(
                          \frac{3\ell^3 \Lambda_0}{2\gamma}
                          - \frac{9\rho^2}{8\ell \gamma}
                       \right) e \,, \nn\\
                        f_{(1)} &\;\to\; F_{(1)} - \frac{\gamma}{6\ell^3}\, e \,.
\end{align}
In terms of these new variables, the dynamical field equations (apart from the torsion constraint, which is unaffected by this change of variables) take the form
\begin{align}
    0 &= R(\omega) + e \times H_{(1)} \,, \nn \\
    0 &= D(\omega) H_{(1)} + e \times F_{(1)} \,, \nn \\
    0 &= D(\omega) F_{(1)}
         + \frac12\, H_{(1)} \times H_{(1)}
         + \frac{\gamma}{3\ell^3}\, e \times H_{(2)} \,.
\end{align}
These imply the following expressions for $H_{(1)}$, $F_{(1)}$ and $H_{(2)}$ \cite{sadik-trancutaion,my-precious}:
\begin{align}
    H_{(1)\mu\nu} &= - S_{\mu\nu} \,, &
    F_{(1)\mu\nu} &= C_{\mu\nu} \,, &
    H_{(2)\mu\nu} &= - \frac{3\ell^3}{\gamma}
       \left( D_{\mu\nu} - P_{\mu\nu} + \frac{1}{4} P\, g_{\mu\nu} \right) \,,
\end{align}
where
\begin{eqnarray}
    D_{\mu\nu} := \epsilon_{\mu}{}^{\rho\sigma} \nabla_{\rho} C_{\nu\sigma} \,,
    \qquad
    P_{\mu\nu} := G_{\mu}{}^{\lambda} S_{\nu\lambda} \,,
\end{eqnarray}
and $P := P_{\mu}{}^{\mu}$. The expressions above are written in terms of the transformed fields. To obtain the solutions for the original auxiliary fields $h_{(1)}$, $f_{(1)}$ and $h_{(2)}$, we now invert the definitions introduced in \eqref{hftoHF}:
\begin{align}
    h_{(1)\mu\nu} &= - S_{\mu\nu}
                    + \frac{\rho}{\ell^2}\, g_{\mu\nu} \,, &
    h_{(2)\mu\nu} &= \frac{3\ell^3}{\gamma}
                    \left( D_{\mu\nu} - P_{\mu\nu}
                           + \frac{1}{4} P\, g_{\mu\nu} \right)
                    - \frac{3\rho \ell}{2\gamma} S_{\mu\nu}
                    - \frac{3}{8\ell \gamma} (2\rho + 1)\, g_{\mu\nu} \,,\nn\\
                     f_{(1)\mu\nu} &= C_{\mu\nu}
                    + \frac{\gamma}{3\ell^3}\, g_{\mu\nu} \,.
\end{align}
The corresponding metric equation is
\begin{eqnarray}
    0 = G_{\mu\nu}
        - \frac{1}{\ell^2} g_{\mu\nu}
        - \epsilon_{\mu}{}^{\rho \sigma} \nabla_{\rho} h_{(2)\nu\sigma}
        + \frac{1}{2}\, \epsilon_{\mu}{}^{\rho\sigma}
                       \epsilon_{\nu}{}^{\lambda\tau}
                       h_{(2)\rho\lambda} h_{(2)\sigma\tau} \,.
\end{eqnarray}
In a maximally symmetric AdS background, the auxiliary fields are proportional to the dreibein, i.e.
\begin{eqnarray} \label{background}
    \omega = \omega(e)\,, \qquad
    h_{(n)} = c_n e\,, \qquad
    f_{(n)} = \beta_n e\,.
\end{eqnarray}
The coefficients $(c_n,\beta_n)$ are model-dependent constants. For the model \eqref{lagrangian}, they are found to be
\begin{eqnarray} \label{bc}
    \{ \Lambda_0, c_1, \beta_1, c_2 \}
    = \left\{
        \frac{(\rho - 1)(1 + 3\rho)}{4\ell^4}\,,\,
        -\frac{\rho - 1}{2\ell^2}\,,\,
        \frac{\gamma}{6\ell^3}\,,\,
        0
      \right\} \,.
\end{eqnarray}
The asymptotic Virasoro $\oplus$ Virasoro symmetry algebra implied by the Brown--Henneaux boundary conditions has central charges that can be computed quite generically in CS-like theories \cite{Merbis:2014vja,charges,on-off}
\begin{eqnarray} \label{central charges}
    c^{\pm}
    = - g^{\text{eff}}_{e\omega}\, c^{\pm}_{\text{GR}}
      + \frac{1}{\ell}\, g^{\text{eff}}_{\omega\omega}\, c^{\pm}_{\text{EG}}
    = \left(
         - g^{\text{eff}}_{e\omega}
         \pm \frac{1}{\ell}\, g^{\text{eff}}_{\omega\omega}
      \right) \frac{3\ell}{2G}\,,
\end{eqnarray}
where
\begin{align} \label{g-eff}
    g^{\text{eff}}_{e \omega}=& g_{e \omega} + c_n g_{\omega h_{(n)}} + \beta_n g_{\omega f_{(n)}}, \nn \\
    g^{\text{eff}}_{\omega \omega }=&g_{\omega \omega }, \nn \\
     g^{\text{eff}}_{e e}=&  g_{ee} +  2c_n g_{eh_{(n)}} + 2 \beta_n g_{ef_{(n)}} + c_n c_m g_{h_{(n)}h_{(m)}} + 2c_n \beta_m g_{h_{(n)}f_{(m)}} + \beta_n \beta_m g_{f_{(n)}f_{(m)}}\,.
\end{align}
In our $N=2$ case, the non-vanishing flavour-metric components are
\begin{equation}
    g_{\omega\omega} = \frac{\gamma}{3\ell}\,, \qquad
    g_{h_{(1)}\omega} = 1\,, \qquad
    g_{h_{(2)}\omega} = - \frac{\gamma}{3\ell}\,.
\end{equation}
Using the background values \eqref{bc}, the central charges of the model become
\begin{eqnarray} \label{central charges-generic}
    c^{\pm}
    = \frac{3}{4\ell G}
      \left( \rho - 1 \pm \frac{2\gamma}{3} \right)\,.
\end{eqnarray}
The region that leads to positive norm of boundary states is determined by the condition
\begin{eqnarray} \label{unitarity}
   c^+ c^- > 0 
   \quad \Rightarrow \quad
   9(\rho - 1)^2 - 4\gamma^2 > 0\,.
\end{eqnarray}
We now look for maximally symmetric solutions
\begin{eqnarray}
    R(\bar{\omega}) + \frac{1}{2\ell^2}\, \bar{e} \times \bar{e} = 0\,,
\end{eqnarray}
and study perturbations about an AdS background, i.e.
\begin{eqnarray} \label{lin}
    e = \bar{e} + k\,, \qquad
    \omega = \bar{\omega} + v\,, \qquad
    h_{(n)} = c_n (\bar{e} + k) + p_n\,, \qquad
    f_{(n)} = \beta_n (\bar{e} + k) + q_n\,,
\end{eqnarray}
where $(k, v, p_n, q_n)$ are perturbations, $(\bar{e}, \bar{\omega})$ are background fields, and the background coefficients $(c_n,\beta_n)$ are given in \eqref{bc}. The second-order (quadratic) Lagrangian for these perturbations is
\begin{align}
    \mathcal{L}^{(2)}  = & 
      \frac{\gamma}{6\ell^3}\, k \cdot \bar{D}k
    + q_1 \cdot \bar{D}k
    + p_1 \cdot \bar{D}v
    + \frac{1}{2\ell^2}(1 - \rho)\, k \cdot \bar{D}v
    + \frac{\gamma}{6\ell}\, v \cdot \bar{D}v
    - \frac{\gamma}{3\ell}\, p_2 \cdot \bar{D}v
     \nn \\
    & + \frac{\gamma}{6\ell}\, p_2 \cdot \bar{D}p_2  + \frac{\gamma}{3\ell^3}\, \bar{e} \cdot v \times k
    + \bar{e} \cdot v \times q_1
    + \frac{1}{4\ell^2}(1 - \rho)\, \bar{e} \cdot v \times v
    + \frac{1}{2}\, \bar{e} \cdot p_1 \times p_1
    \nn \\
    &  - \frac{\gamma}{3\ell^3}\, \bar{e} \cdot k \times p_2  + \frac{1}{\ell^2}\, \bar{e} \cdot k \times p_1
    + \frac{1}{4\ell^4}(1 - \rho)\, \bar{e} \cdot k \times k\,.
\end{align}
The corresponding (decoupled) linearized equations of motion are
\begin{align}
    0 &= \bar{D}k + \bar{e} \times v\,, \nn \\
    0 &= \bar{D}v + \frac{1}{\ell^2}\, \bar{e} \times k + \bar{e} \times p_1\,, \nn \\
    0 &= \bar{D}p_1 + \bar{e} \times q_1\,, \nn \\
    0 &= \bar{D}q_1
         + \frac{1}{2\ell^2}(1 + \rho)\, \bar{e} \times p_1
         - \frac{\gamma}{3\ell^3}\, \bar{e} \times p_2\,, \nn \\
    0 &= \bar{D}p_2 + \bar{e} \times p_1\,. 
\end{align}
The roots of the characteristic polynomial of the linearized system determine the masses of the propagating modes. Introducing $u = m\ell$, where $m$ denotes the eigenvalues of the mass matrix, one finds
\begin{eqnarray} \label{charpol}
    0 = (u^2 - 1)\left( u^3 - \frac{1}{2} u (1 + \rho) + \frac{1}{3}\gamma \right)\,.
\end{eqnarray}
The factor $(u^2 - 1)$ corresponds to two massless modes, while the cubic factor describes three massive modes. To obtain three distinct massive states, the discriminant of the cubic in \eqref{charpol} must be positive. If the discriminant vanishes, two of the massive eigenvalues coincide, yielding a pair of degenerate massive states. In the chiral limit, one of the massive eigenvalues becomes equal to $u = 1$ (i.e.\ $m = 1/\ell$), so that this root is degenerate with algebraic multiplicity two but geometric multiplicity one, admitting a non-diagonal Jordan form. This implies that chiral MMG-like models possess logarithmic modes in their spectrum \cite{Grumiller_2008}. In the following, we examine these cases separately.

\subsection{Models with three distinct massive states}
To ensure that the roots of the cubic equation are real and distinct, the following condition must be satisfied:
\begin{eqnarray}
    \Delta =\frac{1}{2}( (1+\rho)^3 -  6\gamma^2  )>0\,.
\end{eqnarray}
The roots of the depressed cubic equation \eqref{charpol} are found by using Cardano's method
\begin{eqnarray} \label{u}
    u_k:=\frac{\sqrt{6}}{3} \sqrt{1+\rho}\,\cos\left(\frac{1}{3} \cos^{-1}\left(-\frac{\gamma}{1+\rho}\sqrt{\frac{6}{1+\rho}} \right) - \frac{2\pi (k-1)}{3} \right)   \,, \quad k=1,2,3.
\end{eqnarray}
In an AdS background, no-tachyon condition becomes $m_k^2-1/\ell^2\geq0$, which reduces to $u^2\geq 1$. For the solutions in \eqref{u}, no tachyon condition becomes
\begin{eqnarray} \label{no-tachyon}
    u_k^2=\frac{2+2\rho}{3}\cos^2\left(\theta - \frac{2\pi (k-1)}{3} \right)  &\geq &1\,, \quad k=1,2,3.
\end{eqnarray}
Since the no-tachyon condition \eqref{no-tachyon} ensures that $u_k^2-1 \geq 0$ for all $k$, it is convenient to define a set of non-negative quantities 
$r_k:=u_k^2-1$. Accordingly, the elementary symmetric polynomials constructed from $r_k$ must also be non-negative, i.e.
\begin{align}
 e_1(r_1,r_2,r_3):=r_1+r_2+r_3 \geq 0\,, \quad  &  e_2(r_1,r_2,r_3):=r_1r_2+r_1r_3+r_2r_3 \geq 0\,, \nn\\
 & e_3(r_1,r_2,r_3):=r_1r_2r_3 \geq 0\,. \
\end{align}
To evaluate these expressions, the definitions of the elementary symmetric polynomials are used. The coefficients of the cubic equation \eqref{charpol} are then related to the power sums of its roots 
$u_k$ through Newton’s identities. By substituting these relations, one obtains explicit inequalities among the parameters $\rho$ and $\gamma$, which are presented below
\begin{eqnarray} \label{no-tachyon,sympol}
    e_1(r_1,r_2,r_3)&=&    \sum_{i=1}^3 (u_i^2-1) = \rho-2 \geq 0\,, \nn \\
    e_2(r_1,r_2,r_3)&=& \frac{1}{2} \left[ \left(\sum_{i=1}^3 r_i \right)^2 - \sum_{i=1}^3 r_i^2 \right]=\frac{1}{2} \left[ (\rho-2)^2 - \sum_{i=1}^3 (u_i^2-1)^2 \right] =(\rho-5)(\rho-1) \geq 0 \,, \nn \\
    e_3(r_1,r_2,r_3)&=&r_1r_2r_3= \prod_{i=1}^3 \left( u_i^2 -1 \right)= \frac{4\gamma^2}{9}-(1 +\rho)^2 \geq 0\,.
\end{eqnarray}

\begin{figure}
\centering
\begin{tikzpicture}
\begin{axis}[
    axis lines=middle,
    axis on top,
    xlabel={$\rho$},
    ylabel={$\gamma$},
    xmin=-1, xmax=10,
    ymin=-20, ymax=20,
    grid=both,
    width=10cm,
    height=7cm,
    ticklabel style={font=\small},
    label style={font=\small}
]

\path[name path=r5L] (axis cs:5,-20) -- (axis cs:5,20);
\path[name path=r5R] (axis cs:10,-20) -- (axis cs:10,20);

\addplot[draw=none, fill=black!12] fill between[of=r5L and r5R];
\draw[very thick] (axis cs:5,-20) -- (axis cs:5,20);

\addplot[name path=top, draw=none]    coordinates {(-1,20) (10,20)};
\addplot[name path=bottom, draw=none] coordinates {(-1,-20) (10,-20)};

\addplot[name path=upA, domain=-1:10, samples=200, thick]
  {1.5*abs(1 + x)};
\addplot[name path=loA, domain=-1:10, samples=200, thick]
  {-1.5*abs(1 + x)};

\addplot[
  draw=none,
  pattern=north east lines,
  pattern color=black!40,
] fill between[of=upA and top, soft clip={domain=-1:10}];

\addplot[
  draw=none,
  pattern=north east lines,
  pattern color=black!40,
] fill between[of=bottom and loA, soft clip={domain=-1:10}];

\addplot[name path=upB, domain=-1:10, samples=200, thick, dashed]
  {1.5*abs(x - 1)};
\addplot[name path=loB, domain=-1:10, samples=200, thick, dashed]
  {-1.5*abs(x - 1)};

\addplot[draw=none, fill=black!6]
  fill between[of=upB and loB, soft clip={domain=-1:10}];

\addplot[only marks, mark=*] coordinates {(5,6)};
\node[right] at (axis cs:5,6) {$(5,6)$};

\end{axis}
\end{tikzpicture}
\label{parameter space}

\caption{
Shaded regions in the $(\rho,\gamma)$ plane:
light gray strip shows $\rho\ge5$ which is the common region for $\rho$ satisfied by $e_1(r)$ and $e_2(r)$ in \eqref{no-tachyon};
the lightly hatched outer wedge shows $4\gamma^2-9(1+\rho)^2\ge0$;
the dashed inner wedge corresponds to $9(\rho-1)^2-4\gamma^2>0$. The dashed lines correspond to the chiral case, which is discussed in section \ref{chiralsection}. The point $(5,6)$ is a special point discussed in subsection \ref{ultrachiral}. No points satisfy all conditions \eqref{unitarity} and \eqref{no-tachyon} simultaneously. 
} 
\end{figure}
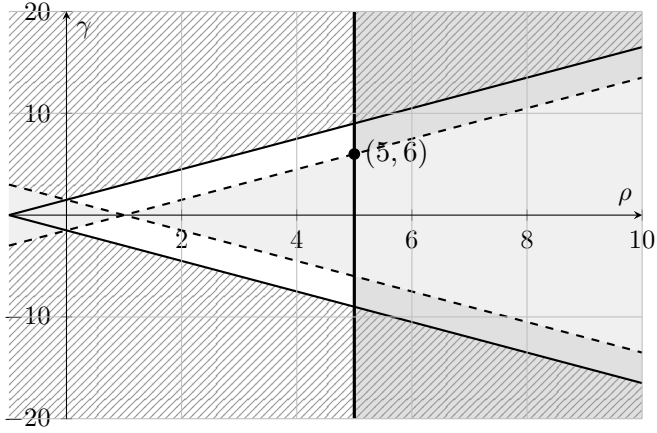

As can be seen in Figure \ref{parameter space}, there is no region in $\rho \gamma$-plane in which both the unitarity condition \eqref{unitarity} and the no-tachyon conditions \eqref{no-tachyon} hold.
To check the no-ghost condition, we need to express the second-order Lagrangian in diagonal form.
For starters, one can write
\begin{eqnarray} \label{lag-x}
    \mathcal{L}^{(2)}=\frac{1}{2}\mathbf{x}^T \cdot \mathcal{K} \cdot \bar{D}\mathbf{x} +\frac{1}{2} \bar{e}  \mathbf{x}^T \cdot \mathcal{P} \cdot \mathbf{x}, \quad \text{where} \quad \mathbf{x}^T:=\begin{pmatrix}
        k & v & p_1 & q_1 & p_2
    \end{pmatrix}\,,
\end{eqnarray}
and $\mathcal{K}$ and $\mathcal{P}$ are $5\times5$ coefficient matrices. The corresponding field equations are
\begin{eqnarray}
    \delta \mathbf{x}^T:  \mathcal{K}  \cdot \bar{D}\mathbf{x} + \bar{e} \mathcal{P} \cdot \mathbf{x} = 0 \qquad \text{or} \qquad
    \bar{D}\mathbf{x} + \bar{e} \mathcal{K}^{-1} \cdot \mathcal{P} \cdot \mathbf{x} =0\,.
\end{eqnarray}
Consider the transformation $\mathbf{y}=S\mathbf{x}$. In terms of the new variables, the second-order Lagrangian becomes
\begin{align} \label{lagy}
     \mathcal{L}^{(2)}=A(\mathbf{y}^T \cdot \bar{D}\mathbf{y} +  \bar{e}   \mathbf{y}^T \cdot M \cdot \mathbf{y}), & \quad \text{with} \quad \mathbf{y}^T:=\begin{pmatrix}
        \psi_0 & \phi_0 & \psi_1 & \phi_1 & \psi_2
    \end{pmatrix}\,, \nn \\
    \delta \mathbf{y}^T \, : \,&\bar{D}\mathbf{y} +  \bar{e} M \cdot \mathbf{y} =0\,,
\end{align} 
   where $M$ is a diagonal matrix $M={\rm diag} (-1/\ell,1/\ell,u_1/\ell,u_2/\ell,u_3/\ell)$, and $S$ is the similarity matrix. If the overall coefficient $A$ is positive, the model does not contain ghost states \cite{MMG,Ozkan-exotic}. 
   
 To obtain the linearized Lagrangian in the diagonal basis, we consider the following transformation
\begin{align}
    \psi_0=&-\frac{1}{2\ell}k +\frac{1}{2} v + \frac{1}{(1+u_1)(1+u_2)(1+u_3)}\left( (1+u_1+u_2+u_3)\ell p_1+\ell^2q_1+u_1u_2u_3 p_2 \right)\,, \nn \\
    \phi_0=&\frac{1}{2\ell}k+\frac{1}{2}v +\frac{1}{(1-u_1)(1-u_2)(1-u_3)}\left((1-u_1-u_2-u_3)\ell p_1+\ell^2q_1+u_1u_2u_3p_2 \right)\,, \nn \\
    \psi_1=&\frac{1}{(u_1-u_2)(u_1-u_3)} (-\ell(u_2+u_3)p_1+\ell^2q_1+u_2u_3p_2)\,, \nn \\
    \phi_1=&\frac{1}{(u_1-u_2)(u_2-u_3)} (\ell(u_1+u_3)p_1-\ell^2q_1-u_1u_3p_2)\,, \nn \\
    \psi_2=&\frac{1}{(u_1-u_3)(u_2-u_3)}(-\ell(u_1+u_2)p_1+\ell^2q_1+u_1u_2p_2)\,,
\end{align}
which gives rise to the second-order Lagrangian in this new basis as
\begin{align}
    \mathcal{L}^{(2)}_{\Delta >0}=&\ell c^+\left(\psi_0 \cdot \bar{D}\psi_0 -\frac{1}{\ell} \bar{e} \cdot \psi_0 \times \psi_0 \right) -\ell c^- \left(\phi_0 \cdot \bar{D}\phi_0 +\frac{1}{\ell} \bar{e} \cdot \phi_0 \times \phi_0 \right)  \nn \\
    & + \kappa_1 \left(\psi_1 \cdot \bar{D}\psi_1 +\frac{u_1}{\ell} \bar{e} \cdot \psi_1 \times \psi_1 \right)+ \kappa_2 \left(\phi_1 \cdot \bar{D}\phi_1 +\frac{u_2}{\ell} \bar{e} \cdot \phi_1 \times \phi_1 \right) \nn \\
    & + \kappa_3
\left(\psi_2 \cdot \bar{D}\psi_2 +\frac{u_3}{\ell} \bar{e} \cdot \psi_2 \times \psi_2 \right)\,,
\end{align}
where
\begin{align}
    \kappa_1=&\frac{u_1(2u_1+u_2)(u_1-u_2)}{2\ell(u_1^2-1)}\,,& \kappa_2=&\frac{u_2(2u_2+u_1)(u_2-u_1)}{2\ell(u_2^2-1)}\,, \nn \\
    \kappa_3=&-\frac{2(u_1^3+u_2^3)+7u_1u_2(u_1+u_2)}{2\ell((u_1+u_2)^2-1)}\,.
\end{align}
As evident from these equations, it is not possible to eliminate ghost states completely, since the positivity condition $\kappa_i>0$ cannot be satisfied simultaneously for all $i$. 
An MMG-like model with three distinct massive states can be ghost-free only at the price of introducing a tachyonic mode. If tachyons are excluded, the theory necessarily becomes non-unitary. More generally, irrespective of the detailed unitarity or tachyon analysis, the structure of the cubic mass polynomial guarantees that at least one of the massive modes is a ghost. 
\subsection{Chiral case} \label{chiralsection}
The condition $c^-=0$ corresponds to $2\gamma=3(\rho-1)$. Along this line, the background solutions are
\begin{eqnarray}
    \{ \Lambda_0, c_1, \beta_1, c_2 \} = \left\{ \frac{(\rho-1)(1+3\rho)}{4\ell^4},\frac{1-\rho}{2\ell^2}, \frac{\rho-1}{4\ell^3},0 \right\} \,.
    \end{eqnarray}
    The central charges of the model are then $c^+ = (\rho - 1)/\ell^2$ and $c^- = 0$, indicating that the theory is chiral. To investigate the presence of tachyonic excitations, one must study the linearized equations of motion, whose quadratic Lagrangian can be written in the form \eqref{lagy} with $M = M_+$. The $5 \times 5$ mass matrix $M_+$ has eigenvalues $\{-1{/\ell}\,,1/{\ell}\,,1/{\ell}\,, m_-/{\ell}\,, m_+/{\ell} \}$, where
    \begin{equation}
        m_{\pm}=\frac{-1\pm\sqrt{2\rho -1}}{2\ell} \,.
    \end{equation}
    Because the eigenvalue $1/\ell$ has algebraic multiplicity two but geometric multiplicity one, the matrix $M_+$ is not diagonalizable. Instead, we bring it to Jordan normal form by constructing an appropriate chain of generalized eigenvectors, as detailed in Appendix~\ref{JSchiral}. No-tachyon condition $m_{\pm}^2 \geq \ell^2$ for the massive states becomes 
 \begin{eqnarray} \label{no tachyon-chiral}
     \rho \geq 5\,.
 \end{eqnarray}
Consider the transformation
\begin{align}
    \psi_0=&\frac{1}{2\ell}\left( -k + \ell v +\frac{\ell^2}{\rho-1} p_1 -\frac{\ell^3}{\rho-1}q_1 -\frac{\ell}{2}p_2 \right) \,, \nn \\
    \phi_0=&\frac{1}{2\ell}(k+\ell v)+\frac{1}{4(\rho-5)^2}\left(2\ell(13-5\rho)p_1-6\ell^2(\rho-1)q_1+(\rho-1)(7+\rho)p_2\right
    )\,, \nn \\
    \psi_1=&\frac{1}{5-\rho}\left(p_1+\ell q_1+\frac{\rho-1}{2\ell}p_2 \right) \,, \nn \\
    \phi_1=&\frac{1}{\Upsilon(\rho-5)} \left(\ell(2-\rho+\Upsilon) p_1+\ell^2(3-\Upsilon)q_1 +(1+\rho-2\Upsilon)p_2 \right)\,, \nn \\
    \psi_2=&\frac{1}{\Upsilon(\rho-5)}\left(\ell(-2+\rho+\Upsilon) p_1-\ell^2(3+\Upsilon)q_1+(1+\rho-2\Upsilon)p_2 \right)\,,
\end{align}
where $\Upsilon:=\sqrt{2\rho-1}$.
The second-order Lagrangian in this new basis is
\begin{align}
     \mathcal{L}^{(2)}_c =&\ell c^+ \psi_0 \cdot \left( \bar{D}\psi_0 -\frac{1}{\ell} \bar{e}  \times \psi_0 \right) +(5-\rho)\psi_1 \cdot \left( \bar{D}\phi_0 +\frac{1}{\ell} \bar{e}  \times \phi_0 +\frac{1}{2} \bar{e} \times \psi_1 \right) \nn \\
    & + \frac{\ell}{4}(5\rho-13)\psi_1 \cdot \left( \bar{D}\psi_1 +\frac{1}{\ell} \bar{e}  \times \psi_1 \right) + \kappa_+ \phi_1 \cdot\left( \bar{D}\phi_1 +m_- \bar{e}  \times \phi_1 \right) \nn \\
    &  + \kappa_-\psi_2 \cdot
\left( \bar{D}\psi_2 +m_+ \bar{e}  \times \psi_2 \right)\,,
\label{L2c}
\end{align}
where
\begin{eqnarray}
    \kappa_{\pm} := \frac{(\rho-1)\left(3(1-2\rho)\mp (4+\rho)\Upsilon\right)}{2\ell \left(\Upsilon\pm(\rho-2)\right)^2}\,.
\end{eqnarray}
There is no $\rho$ which makes both $\kappa_+$ and $\kappa_-$ positive. Therefore, the presence of a ghost state cannot be avoided. The second term on the right hand side of \eqref{L2c} is negative
because of the no-tachyon condition \eqref{no tachyon-chiral}. This term and the one after that 
describe the presence of a Jordan chain, which is in fact expected \cite{Grumiller_2008}. The appearance of a Jordan block signals that the chiral limit of the MMG-like model corresponds to a logarithmic CFT.

\subsection{Models with degenerate massive states} \label{degMassStates}

We now consider the case in which the cubic equation \eqref{charpol} develops multiple roots. This occurs when its discriminant vanishes:
\begin{eqnarray}
    \Delta= \frac{1}{2} (1+\rho)^3 - 3 \gamma^2&=&0 \,, \qquad \text{with} \qquad \gamma=\pm\frac{(1+\rho)^{3/2}}{\sqrt{6}}\,.
\end{eqnarray}
Let us focus on the case of positive $\gamma$. The central charges of the model are then
\begin{eqnarray} \label{central charges-degmass}
    c^{\pm}=\frac{1}{12G\ell^2}\left(9(\rho-1)\pm\sqrt{6}(1+\rho)^{3/2}\right)\,.
\end{eqnarray}
In this branch, the unitarity condition \eqref{unitarity} reduces to
\begin{eqnarray} \label{unitarity-degenerate mass}
    -1<\rho<\frac{1}{2}\,.
\end{eqnarray}
The roots of the cubic equation are
\begin{eqnarray} \label{degenerate masses}
    u_1=-\frac{\sqrt{6}}{3}(1+\rho)^{1 /2}=-2u\,, \qquad u_2=u_3=\frac{\sqrt{6}}{6}(1+\rho)^{1/ 2}=:u\,.
\end{eqnarray}
No-tachyon condition implies that $u_k^2\geq 1$, thus
\begin{eqnarray}
    u_1^2=\frac{2}{3}(1+\rho)\geq 1 \Rightarrow \rho \geq \frac{1}{2}\,, \qquad u_2^2=u_3^2=\frac{1}{6}(1+\rho)\geq 1 \Rightarrow \rho \geq 5\,,
\end{eqnarray}
which clearly contradicts the unitarity window \eqref{unitarity-degenerate mass}. Thus, in the degenerate-mass case there is no parameter region in which both unitarity and the absence of tachyons are simultaneously satisfied. The eigenvalues of the mass matrix $M$ appearing in \eqref{lagy} are $\left\{ -1/{\ell}\,, 1/{\ell}\,, u/{\ell}\,, u/{\ell}\,,-2u/{\ell}\right\}$. Since there are two degenerate massive modes with the same eigenvalue $u/\ell$, the mass matrix is not diagonalizable. Instead, it admits a Jordan 2-block, whose explicit construction is given in Appendix~\ref{JS-degmass}. To analyze the no-ghost condition, we perform the following field redefinitions:
\begin{align}
    \psi_0=&-\frac{1}{2\ell}k+\frac{1}{2} v+ \frac{1}{2(1+u)^2(2u-1)}(\ell p_1-\ell^2 q_1-2u^3p_2)\,, \nn \\
    \phi_0=&\frac{1}{2\ell}k+\frac{1}{2} v+\frac{1}{2(u-1)^2(1+2u)}(\ell p_1+\ell^2q_1-2u^3p_2)\,, \nn \\
    \psi_1=&\frac{2}{9(1-4u^2)} \left(-2p_1+\frac{\ell}{u}q_1+\frac{u}{\ell}p_2 \right)\,, \nn \\
    \phi_1=&\frac{1}{9(u^2-1)^2} \left(-2(1+2u^2)p_1+\frac{\ell}{u}(1-7u^2)q_1+\frac{4u}{\ell}(5u^2-2)p_2 \right)\,, \nn \\
    \psi_2=&\frac{1}{3\ell^2(1-u^2)} (\ell u p_1 -\ell^2 q_1+2u^2p_2)\,.
\end{align}
In this new basis, the quadratic Lagrangian takes the following form
\begin{align}
    \mathcal{L}^{(2)}_{\Delta=0} =&\ell c^+ \psi_0 \cdot \left(\bar{D}\psi_0-\frac{1}{\ell} \bar{e} \times \psi_0 \right)-\ell c^- \phi_0 \cdot \left(\bar{D}\phi_0+\frac{1}{\ell} \bar{e} \times \phi_0 \right)\,,\nn\\
    &+\frac{9}{4}u \ell (1-4u^2) \psi_1 \cdot \left(\bar{D}\psi_1+\frac{2u}{\ell} \bar{e} \times \psi_1 \right)  +\frac{\ell^3}{4}(5u^2-2) \psi_2 \cdot \left( \bar{D}\psi_2 +\frac{u}{\ell} \bar{e} \times \psi_2 \right) \nn \\
    & +\frac{3\ell^2}{2}(u^2-1)\psi_2 \cdot \left(\bar{D} \phi_1 + \frac{u}{\ell} \bar{e} \times \phi_1 +\frac{1}{2} \bar{e} \times \psi_2 \right)\,.
\end{align}
In the absence of tachyons, i.e.\ for $u^2 \geq 1$, the factor $(1 - 4u^2)$ is negative, so the massive mode with mass $2u$ is therefore a ghost. By contrast, the pair of degenerate massive modes with mass $u$ has positive energy. 

\subsubsection{Chiral case} \label{ultrachiral}

Let $c^-$ in \eqref{central charges-degmass} vanish. This condition fixes
 $(\rho,\gamma)=(5,6)$ which corresponds to the special case $u=1$ in \eqref{degenerate masses}. The background solutions are then
\begin{eqnarray}
    \left \{ \Lambda_0, c_1, \beta_1, c_2 \right\} = \left\{ \frac{16}{\ell^4},-\frac{2}{\ell^2},\frac{1}{\ell^3},0 \right\}\,. 
\end{eqnarray}
To examine the absence of tachyonic and ghost excitations, we turn to the linearized equations of motion. The mass matrix appearing in \eqref{lagy} will be denoted by $M_c$ in this chiral case, to distinguish it from a generic mass matrix $M$. Its eigenvalues are $\{-2/{\ell}\,,-1/{\ell}\,,1/{\ell}\,,1/{\ell}\,,1/{\ell}\}$. The Jordan normal form of $M_c$ and the corresponding similarity transformation are presented in Appendix~\ref{JS-ultrachiral}. Using this similarity transformation, we introduce a new basis of fields
\begin{align}
     \psi_0 =& \frac{1}{9} (-2\ell p_1+\ell^2q_1+p_2)\,, & \phi_0 =& \frac{1}{8\ell}(-4k+4\ell v +\ell^2p_1-\ell^3q_1-2\ell p_2)\,, \nn \\
    \psi_1=&\frac{1}{216\ell}(108k+108\ell v+37\ell^2p_1-5\ell^3q_1+22\ell p_2)\,, &\phi_1=&\frac{1}{36\ell}(13\ell p_1+7\ell^2q_1-2p_2) \,, \nn \\
    \psi_2=&\frac{1}{6\ell^2}(\ell p_1+\ell^2 q_1-2p_2)\,.
\end{align}
In terms of these redefined fields, the quadratic Lagrangian becomes
\begin{align} \label{ultralog-second-order-lag}
   \mathcal{L}^{(2)}_{\text{ultra-log}}=&-\frac{3}{\ell} \psi_0 \cdot \left(  \bar{D}\psi_0-\frac{2}{\ell} \bar{e}  \times \psi_0 \right) +\frac{4}{\ell}\phi_0 \cdot \left(\bar{D}\phi_0-\frac{1}{\ell}\bar{e} \times \phi_0 \right)  \nn \\
    & +3\ell \phi_1 \cdot \left(\bar{D}\phi_1+\frac{1}{\ell} \bar{e} \times \phi_1 +2\bar{e} \times \psi_2 \right)  + 6\ell \psi_1 \cdot \left(\bar{D}\psi_2 +\frac{1}{\ell} \bar{e} \times \psi_2 \right)  \nn \\
    & +11\ell^3 \psi_2 \cdot \left(\bar{D}\psi_2 +\frac{1}{\ell} \bar{e} \times \psi_2 \right) -13\ell^2 \psi_2 \cdot \left(\bar{D}\phi_1+\frac{1}{\ell} \bar{e} \times \phi_1 +\frac{1}{2}\bar{e} \times \psi_2 \right)  \,.
\end{align}
From \eqref{ultralog-second-order-lag} it is clear that the massive mode is a ghost. Moreover, the rank-3 Jordan structure of $M_c$ implies the presence of two logarithmic modes in the spectrum, which motivates the terminology ``ultra-logarithmic’’ for this chiral point. By contrast, TMG and NMG contain fewer fields; their logarithmic limits lead only to a rank-2 Jordan structure. 

 \section{Conclusions} \label{conclusions}
\paragraph{}
In this work we have developed the CS-like construction of the generic MMG-like family of third-way consistent massive gravity models and presented a detailed analysis of the simplest non-trivial example, namely the $N=2$ case. Starting from the torsional $N=2$ Lagrangian, we derived the full set of first-order field equations and solved explicitly for the auxiliary fields. To determine the physical spectrum, we linearized the theory around a maximally symmetric AdS background and constructed the quadratic action. The characteristic polynomial of the corresponding mass operator is of fifth order and factorizes into a quadratic part describing two massless modes and a cubic part whose roots correspond to three massive excitations. When the discriminant of the cubic is positive, the model admits three distinct massive states; we obtained their explicit expressions using Cardano’s formula. However, combining the bulk and boundary consistency conditions shows that the model can satisfy unitarity only at the price of introducing a tachyonic massive mode. If tachyons are excluded, the theory necessarily becomes non-unitary. Moreover, the diagonalized kinetic matrix reveals that at least one of the massive modes is unavoidably a ghost.

We also investigated special regions of parameter space where the spectrum develops degeneracies. When one of the Brown--Henneaux central charges vanishes, one obtains a chiral line parametrized by two independent parameters. Along this line the linearized mass operator becomes non-diagonalizable. We resolved this by constructing an explicit rank-2 Jordan form, and the kinetic matrix obtained via the associated similarity transformation shows that one of the massive modes remains a ghost. A further degeneracy occurs when the discriminant of the cubic vanishes, yielding two degenerate massive states. In this branch, the no-tachyon requirement is incompatible with unitarity, and a linearized analysis shows that the non-degenerate massive mode necessarily carries negative kinetic energy. The chiral limit of this degenerate-mass branch singles out a special point on the chiral line where the degeneracy is enhanced: three states coincide and the Jordan structure has rank~3, producing two logarithmic partners. This motivates the interpretation of this point as an ``ultra-logarithmic'' limit.

Taken together, by combining the background analysis, the mass spectrum, and the consistency conditions, this work provides a comprehensive description of the $N=2$ MMG-like model. In particular, it uncovers new structural features, most notably the emergence of higher-rank Jordan blocks, which provide strong evidence that the corresponding boundary theory is a logarithmic CFT. It would be interesting to further clarify the holographic interpretation of the rank-3 (ultra-logarithmic) point, and to extend the present analysis to higher-$N$ models and to possible supersymmetric or higher-spin generalizations.

\appendix \label{appendix}
\section*{Appendix}
\addcontentsline{toc}{section}{Appendix}

In the transformed basis, the second-order Lagrangian and the corresponding linear equations are given in \eqref{lagy}. In this appendix, we will look at non-diagonalizable mass matrices. In each following subcases, one of the eigenvalues of the mass matrix is degenerate with algebraic multiplicity two (three) but geometric multiplicity one. This allows us to write a Jordan form $J$ which can be obtained via a similarity transformation $S^{-1}\cdot M \cdot S=J$.

\section{Jordan form and similarity matrix in section \ref{chiralsection}} \label{JSchiral}
The chain of the generalized eigenvectors discussed in section \ref{chiralsection} is
\begin{equation}
\begin{aligned} \label{Dv=0}
  \left(M_+ - \frac{1}{\ell} I\right) \mathbf{v_1} &= 0, \quad &
  \left(M_+ + \frac{1}{\ell} I\right) \mathbf{v_2} &= 0, \quad &
  \left(M_+ + \frac{1}{\ell} I\right) \mathbf{v_3} &= \mathbf{v_2}, \\
  \left(M_+ + m_+ I\right) \mathbf{v_4} &= 0, \quad &
  \left(M_+ + m_- I\right) \mathbf{v_5} &= 0,
\end{aligned}
\end{equation}
The Jordan form is obtained via a similarity transformation $S^{-1}_+\cdot M_+ \cdot S_+=J_+$ which are given as
\begin{eqnarray}
   S_+=\begin{pmatrix}
        -\ell & \ell & -\ell^2 & -\frac{\ell(1+\Upsilon)}{\rho-2+\Upsilon} & \frac{\ell(-1+\Upsilon)}{\rho-2-\Upsilon} \\
        1 & 1 & 0 & \frac{\rho+\Upsilon}{\rho-2+\Upsilon} & \frac{-\rho+\Upsilon}{\rho-2+\Upsilon} \\
        0 & 0 & 2 & -\frac{1+\Upsilon}{2\ell} & \frac{-1+\Upsilon}{2\ell} \\
        0 & 0 & \frac{2}{\ell} & \frac{\rho+\Upsilon}{2\ell^2} & \frac{\rho-\Upsilon}{2\ell^2} \\
        0 & 0 & 2\ell & 1 & 1
    \end{pmatrix}\,, \qquad  J_+=\begin{pmatrix}
        -\frac{1}{\ell} & 0 & 0 & 0 & 0 \\
        0 & \frac{1}{\ell} & 1 & 0 & 0 \\
        0 & 0 & \frac{1}{\ell} & 0 & 0 \\
        0 & 0 & 0 & m_- & 0 \\
        0 & 0 & 0 & 0 & m_+
    \end{pmatrix}\,,
\end{eqnarray}
$J_+$ denotes the Jordan form of the chiral model characterized by a single non-vanishing, positive central charge $c^+$. 
\section{Jordan form and similarity matrix in section \ref{degMassStates}}\label{JS-degmass}

The linearized equations of motion of the model in section \ref{degMassStates} have the form 
$\bar{D}{\bf x} + \bar{e} M \cdot  {\bf x}=0$, where $M$ is a $5 \times 5$ matrix with eigenvalues 
$\left\{-\frac{1}{\ell},\frac{1}{\ell},-2u,u,u\right\}$. Since one of the eigenvalues is degenerate with algebraic multiplicity two but geometric multiplicity one, the matrix $M$ cannot be diagonalized. Its Jordan normal form is obtained via the following chain
\begin{equation}
\begin{aligned} \label{Dv=0,c}
  \left(M - \frac{1}{\ell} I\right) \mathbf{w_1} &= 0, \quad &
  \left(M + \frac{1}{\ell} I\right) \mathbf{w_2} &= 0, \quad &
  \left(M -2u I\right) \mathbf{w_3} &= 0, \\
  \left(M + u I\right) \mathbf{w_4} &= 0, \quad &
  \left(M + u I\right) \mathbf{w_5} &= \mathbf{w_4},
\end{aligned}
\end{equation}
The similarity transformation that gives the Jordan form is $S^{-1}\cdot M \cdot S=J$,
\begin{eqnarray} \label{SJ-deg.mass}
    S=\begin{pmatrix}
        -\ell & \ell & \ell^2 & \ell^2 & \frac{\ell^3}{u} \\
        1 & 1 & -2u\ell & u\ell & 2\ell^2 \\
        0 & 0 & 4u^2-1 & u^2-1 & 3u\ell -\frac{\ell}{u} \\
        0 & 0 & -\frac{2u}{\ell}(4u^2-1) & \frac{u}{\ell}(u^2-1) &4u^2-2 \\
        0 & 0 & -\frac{\ell}{u}(4u^2-1) & \frac{\ell}{u}(u^2-1) & 2\ell^2
    \end{pmatrix}\,, \qquad J=\begin{pmatrix}
        -\frac{1}{\ell} & 0 & 0 & 0 & 0 \\
        0 & \frac{1}{\ell} & 0 & 0 & 0 \\
        0 & 0 & -2u & 0 & 0 \\
        0 & 0 & 0 & u & 1 \\
        0 & 0 & 0 & 0 & u
    \end{pmatrix}\,.  
\end{eqnarray}
\section{Jordan form and similarity matrix in subsection \ref{ultrachiral}} \label{JS-ultrachiral}
 We construct the Jordan normal form of the mass matrix of the model in subsection \ref{ultrachiral} by introducing the following chain of generalized eigenvectors:
\begin{equation}
\begin{aligned}
  \left(M_c - \frac{2}{\ell} I\right) \mathbf{u_1} &= 0, \quad &
  \left(M_c - \frac{1}{\ell} I\right) \mathbf{u_2} &= 0, \quad &
  \left(M_c + \frac{1}{\ell} I\right) \mathbf{u_3} &= 0, \\
  \left(M_c + \frac{1}{\ell} I\right) \mathbf{u_4} &= \mathbf{u_3}, \quad &
  \left(M_c + \frac{1}{\ell} I\right) \mathbf{u_5} &= \mathbf{u_4}.
\end{aligned}
\end{equation}
The Jordan normal form is obtained via a similarity transformation $S_c^{-1}\cdot M_c \cdot S_c=J_c$ where the similarity matrix $S_c$ and the corresponding Jordan normal form $J_c$ are
\begin{eqnarray}
    S_c=\begin{pmatrix}
        -\frac{2\ell}{3} & -\ell & \ell & -\ell^2 & \ell^3 \\
        \frac{4}{3} & 1 & 1 & 0 & 0 \\
        -\frac{2}{\ell} & 0 & 0 & 2 & -\ell \\
        \frac{4}{\ell^2} & 0 & 0 & \frac{2}{\ell} & 1 \\
        1 & 0 & 0 & 2\ell & -3\ell^2
    \end{pmatrix}\,, \qquad 
    J_c=\begin{pmatrix} \label{j}
        -\frac{2}{\ell} & 0 & 0 & 0 & 0 \\
        0 & -\frac{1}{\ell} & 0 & 0 & 0 \\
        0 & 0 & \frac{1}{\ell} & 1 & 0 \\
        0 & 0 & 0 &\frac{1}{\ell} & 1 \\
        0 & 0 & 0 & 0 & \frac{1}{\ell}
    \end{pmatrix}\,,
\end{eqnarray}
Note that $J_c$ can be obtained from \eqref{SJ-deg.mass} by setting $(\rho,\gamma)=(5,6)$. 

\bibliographystyle{utphys.bst}
\bibliography{reference}

\end{document}